# Cooperative Energy Transfer Controls the Spontaneous Emission Rate Beyond Field Enhancement Limits

Mohamed ElKabbash[1*], Ermanno Miele[1,2], Ahmad K. Fumani[1], Michael S. Wolf[1], Angelo Bozzola[2], Elisha Haber[1], Tigran V. Shahbazyan[3], Jesse Berezovsky[1], Francesco De Angelis[2], Giuseppe Strangi[1,2,4*].

[1.] Department of Physics, Case Western Reserve University, 10600 Euclid Avenue, Cleveland, Ohio 44106, USA.

[2.] IIT - Istituto Italiano di Tecnologia, via Morego 30, 16163 Genova, Italy.

[3.] Department of Physics, Jackson State University, Jackson, Mississippi 39217, USA.

[4.] CNR-NANOTEC, Istituto di Nanotecnologia and Department of Physics, University of Calabria, Italy.

*Correspondence authors E-mail: mke23@case.edu, gxs284@case.edu.

**Quantum emitters located in proximity to a metal nanostructure individually transfer their energy via near-field excitation of surface plasmons. The energy transfer process increases the spontaneous emission (SE) rate due to plasmon-enhanced local field. Here, we demonstrate significant acceleration of quantum emitter SE rate in a plasmonic nano-cavity due to cooperative energy transfer (CET) from plasmon-correlated emitters. Using an integrated plasmonic nano-cavity, we realize up to six-fold enhancement in the emission rate of emitters coupled to the same nano-cavity on top of the plasmonic enhancement of the local density of states. The radiated power spectrum retains the plasmon resonance central frequency and lineshape, with the peak amplitude proportional to the number of excited emitters indicating that the observed cooperative SE is distinct from super-radiance. Plasmon-assisted CET offers unprecedented control over the SE rate and allows to**

**dynamically control the spontaneous emission rate at room temperature enabling an SE rate based optical modulator.**

Ordinary fluorescence arises from the decay of excited quantum emitters (QEs) to lower energy states by spontaneous emission (SE) where the QEs interact independently with the radiation field. The interaction with the radiation field can be controlled by modifying the emitter's electromagnetic environment. In the Wigner-Weisskopf approximation, the SE rate is directly proportional to the electromagnetic local density of states (LDOS) $\rho(\omega, r)$ which characterizes the interaction of a QE, located at $r$, with electromagnetic modes at $r$ [1-3]. The LDOS, thus, represents the number of electromagnetic modes available for the emitter to radiate into per unit volume and frequency interval and it can be modified by, e.g., placing an emitter inside a cavity. The cavity enhanced SE rate is proportional to the ratio of the cavity quality factor $Q$ to the modal volume $V$ ($\propto Q/V$), which is known as the Purcell effect [3]. The emitters' SE rate has been significantly enhanced using plasmonic nanocavities (PNCs) supporting localized surface plasmon (LSP) modes [2-5]. The LDOS enhancement in a PNC results from the strong electromagnetic field confinement within small plasmon mode volume, so a QE transfers its energy to a resonant plasmon mode with an energy transfer rate $\Gamma^{ET}$ that is much faster than the free space SE rate (**Fig. 1a**). Subsequently, a PNC will act as an optical antenna that radiates the transferred energy to the far field with a significantly faster rate due to its large size and dipole moment [2,6]. Accordingly, following the excitation of a QE, the emission rate is proportional to $\Gamma^{ET}$. However, the SE rate of an individual QE is restricted by ultimate limits on plasmonic field enhancement imposed by the Ohmic losses in metals and nonlocal effects near the metal-dielectric interface [7,8].

When an ensemble of QEs is coupled to a plasmonic structure, SE can be greatly accelerated by cooperative effects arising from plasmon-assisted correlations between QEs. For example, interactions of QE with common radiation field enhanced by resonant Mie scattering are predicted to lead to plasmon-enhanced super-radiance characterized by SE at a rate proportional to the *full ensemble size* that includes both excited and ground-state QEs[9-13]. However, the plasmonic enhancement of radiation coupling is offset by relatively strong absorption, compared to scattering, in small metal structures [10], which inhibits coherence buildup that precedes super-radiance burst from incoherently excited emitters [14,15]. An observation of plasmon-enhanced super-radiance remains a challenge as it hinges on delicate interplay between QEs' direct interactions, plasmon-enhanced radiation coupling, and metal losses [16].

On the other hand, it has been recently pointed out that strong plasmon absorption leads to another cooperative effect in a system of *N* excited QEs coupled to a plasmonic resonator that does *not* require coherence buildup between excited QEs [17,18]. If plasmon frequency is tuned to resonance with QEs emission frequency, the indirect plasmonic coupling between QEs gives rise to collective states that transfer their energy to a plasmon *cooperatively* at a rate $\Gamma_c^{ET} = \sum_i^N \Gamma_i^{ET}$ where $\Gamma_i^{ET}$ is the energy transfer rate of individual QEs (**Fig. 1b**). Note that the Förster resonance energy transfer rate from QEs to a plasmon determined by the spectral overlap between the donor (QE) emission band and the acceptor (plasmon) absorption band [19]. Since the plasmon spectral band is normally much broader than that of QEs, the cooperative energy transfer (CET) rate is relatively insensitive, in contrast to super-radiance [20,21], to natural variations of QEs emission frequencies, e.g., due to direct dipole coupling. Following CET from a collective state to the PNC mode, the possible energy flow pathways include (i) energy transfer from PNC to QEs, (ii) energy dissipation within PNC through Ohmic losses, and (iii) PNC antenna radiation. If the antenna's radiation efficiency is sufficiently high, while the overlap between QEs' emission and absorption bands is relatively weak, the energy is mainly radiated away at approximately the same rate, $\Gamma_c^{ET}$, as it is being transferred from QEs. Note that the values of individual rates $\Gamma_i^{ET}$ are determined by the plasmon LDOS at the QEs' positions and can vary significantly depending on the system geometry [17,19]. However, if the LDOS does not change significantly in the region QEs are distributed in, which is the case for PNC we study, the individual QE rates $\Gamma_i^{ET}$ are all comparable and so the cooperative rate $\Gamma_c^{ET}$ scales *linearly with the number of excited emitters*. Since the latter is proportional to the excitation power, the ensemble SE mediated by CET to plasmonic antenna can be controlled directly by the excitation power.

Here, we report the experimental observation of a cooperative SE from an ensemble of *N* excited QEs resonantly coupled to a PNC that acts as a plasmonic antenna. We observe a significant increase of the ensemble SE rate relative to the plasmonic LDOS enhancement, up to six-fold in our samples, which is linear in the excitation power. At the same time, the measured photoluminescence spectrum retains the plasmon resonance lineshape while the overall emission intensity increases linearly with the excitation power. These observations imply that the radiation is emitted by the plasmonic antenna following CET from the excited QEs [18]. The linear dependence of the ensemble SE rate on the number of excited QEs (as opposed to total number of emitters [21-23]) has not, to our knowledge, been observed previously. Such dependence as well as the incoherent nature of CET mechanism [17,18] that does not require coherence buildup [14,15], in contrast to super-radiance, provides a unique possibility for dynamic control of the SE rate in the *same* electromagnetic environment by varying excitation power.

We experimentally demonstrate dynamic control over SE rate by modulating the excitation power, resulting in *reversible* increase and decrease of the SE rate. It is worth to highlight the ability of CET mechanism to dynamically control the SE rate in real time at room temperature, which was only possible in previous works using complex photonic devices at cryogenic temperatures[24,25]. Note that cooperative enhancement of the ensemble SE rate take place on top of the plasmon LDOS enhancement for individual emitter's SE rate, as we show later in this paper, paving the way towards SE rate control beyond the field enhancement limits imposed by losses and nonlocal effects [7,8]. This is particularly important for short-distance optical communication, where augmenting the modulation rate requires SE rate enhancement [6], and for optical data storage [26] that require faster SE rate to increase the data reading speed.

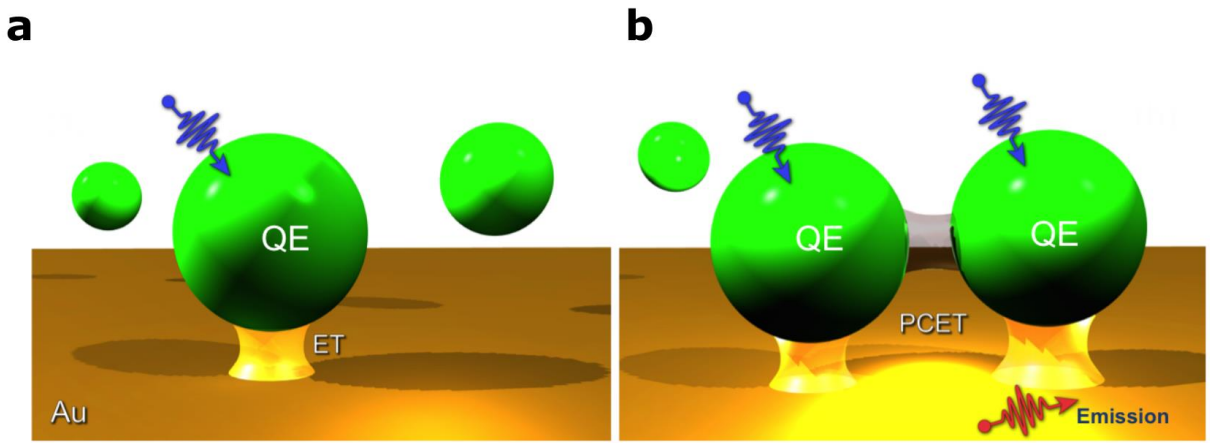

**Figure 1| Schematic of individual and cooperative energy transfer:** **(a)** An excited QE coupled to plasmonic resonator non-radiatively transfers its energy, at a rate $\mathbf{\Gamma}^{ET}$, to the plasmon mode, which radiates it away. **(b)** An ensemble of QEs coupled to a resonant plasmon mode transfer their energy to it cooperatively at a rate $\mathbf{\Gamma}_c^{ET}$ that is the sum of individual rates[17].

To demonstrate the effect, we fabricated three dimensional, out-of-plane hollow PNC [27,28] (*see Supplementary information*). Fig. 2a shows a SEM image of an array of the PNCs. Fig. 2b shows a SEM image of the cross-section of a single PNC that was cut using Focused Ion Beam (FIB). The PNCs are composed of a cylindrical polymeric scaffold, 20 nm thick and 450 nm height, on which a 20 nm gold layer was conformally deposited. The radiation pattern from the out-of-plane PNC is highly directional and the large size of the PNC increases the antenna radiative efficiency [4,6,29]. This is to ensure that the major energy pathway following the energy transfer process is antenna radiation and that the collected photons are mainly from antenna radiation. The height of the nanoantenna was chosen to ensure strong radiation directionality (*Supplementary information,* **Fig.2**). CdSe/ZnS quantum dots (QDs) were spin-coated on the polymeric scaffold onto which the plasmonic shell is formed (**Fig. 2c** and **2d**). We chose QDs as our QEs over, e.g., fluorophores, as they (i) have larger dipole moments which increases the efficiency of non-radiative energy transfer[5], and (ii) exhibit relatively weak absorption in the photoluminescence frequency range to reduce reabsorption which is important to demonstrate CET as we mentioned earlier (*Supplementary information*, **Fig. S3**). The integrated PNC is designed such that QEs are at approximately the same distance away from the plasmonic shell to excite LSPs with the same energy transfer rate, i.e., $\Gamma_c^{ET} \approx N\, \Gamma^{ET}$ (**Fig. 2d**). We note that this relation is robust even for large fluctuations in QEs positions since the LSP electric field inside PNC is nearly uniform. The measured and calculated LSP resonance of the PNC are in close agreement as shown in **Fig. 2e** and **2f**, respectively. To control for other QD-metal interactions that are not related to the excitation of LSPs, we prepared a reference sample where the QDs were spin coated on top of an Au film. **Fig. 2g** compares the QDs photoluminescence collected from a single PNC (red dots) and the photoluminescence from the reference sample (black dots) with excitation wavelength $490\ nm$ and intensity 18.5 W/cm². The photoluminescence maximum is blue shifted from 638 nm (reference) to 631nm (PNC) and is pulled towards the LSP resonance peak (~628nm)[30]. The blue shift in the photoluminescence maximum along with the high directionality and radiative efficiency of our PNC ensure that most of the collected photoluminescence is from the nano-antenna and is due to the excitation of LSPs[29,30].

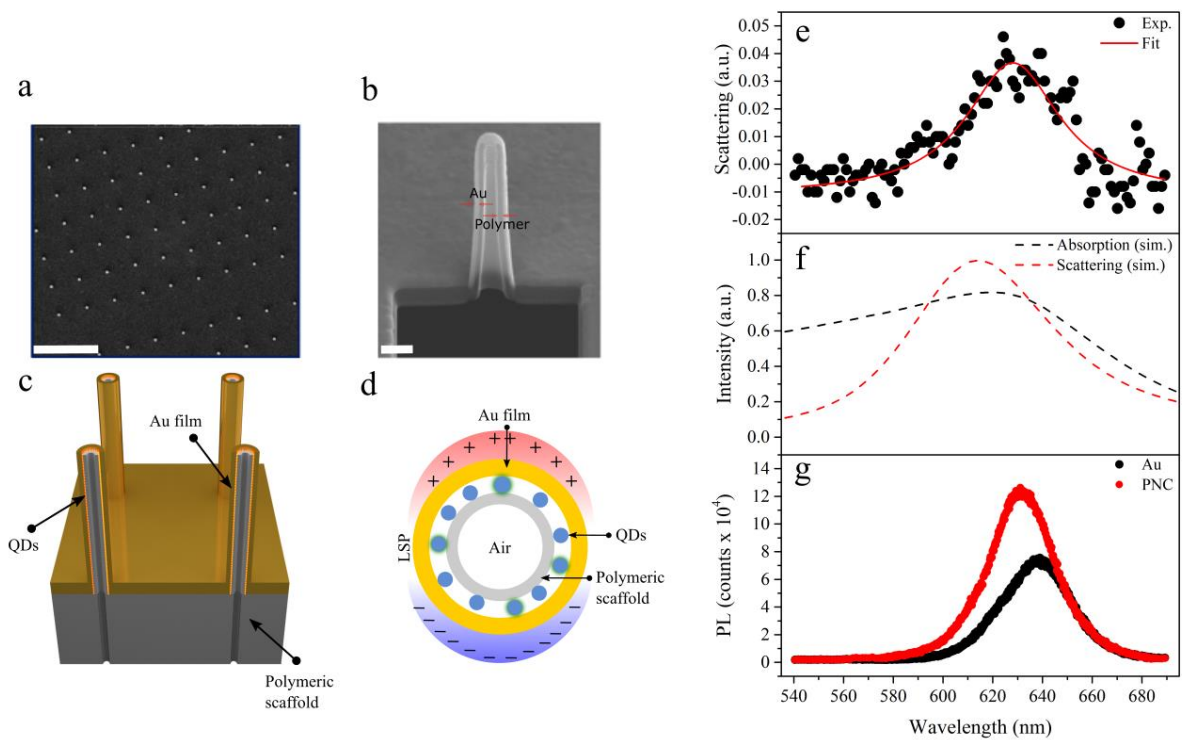

**Figure 2| Characterization of integrated plasmonic nanocavities:** (**a**) SEM image, top view, of plasmonic nanocavity (PNC) array (scale bar= 5 μm). (**b**) an SEM image of a cross-section of a single PNC that was cut using FIB (scale bar = 100 nm). (**c**) a schematic of the nano-pillar PNC. The quantum dots (QDs) are spin-coated on a polymeric scaffold, then an Au layer is deposited. (**d**) Schematic of a cross-section of a single nanopillar. Incident light excites QDs that, subsequently, transfer their energy to excite localized surface plasmons (LSPs) which decay into a photon. (**e**) The measured scattering for PNC array; the resonance maximum was determined by fitting the data with a Lorentzian function. The measured resonance closely agrees with the calculated absorption (black dashed line) and scattering (red dashed line) presented in (**f**). (**g**) Shows the photoluminescence of the QDs spin coated on an Au film (black dots) compared to QD incorporated in a single PNC (red spheres).

Using a confocal laser setup, we were able to locate and measure signals from single PNCs. In particular, the time-resolved emission of the QDs was measured at 630 nm for a range of pump intensities (3.7 W/cm$^2$- 74 W/cm$^2$) and 490 $nm$ excitation wavelength (see *Supplementary information*). **Figure 3a** shows the time-resolved photoluminescence collected from a single PNC (top) and the reference Au film (bottom). The measured photoluminescence lifetime for the reference sample shows no changes upon increasing the pump intensity. On the other hand, the photoluminescence lifetime from the PNC strongly depends on the excitation intensity. We fitted the photoluminescence decay curves with bi-exponential functions obtaining two characteristic decay times: one fast SE rate due to a short living state and a second slow SE rate due to long living states, as shown in **Fig. 3b**. It is known that CdSe/ZnS quantum dots have fast and slow SE rate components (*Supplementary information*, **Fig. S4**)[31]. Strikingly, by increasing the pump intensity, the SE rates increased linearly up to six-fold for the PNCs, while no changes were measured for the Au film, as shown in Fig. 3b. This linear dependence of the SE rate on the excitation intensity, accompanied by a linear increase of the photoluminescence is a clear signature of a plasmon mediated collective energy transfer effect, as thoroughly discussed below.

The demonstrated dynamic control of the SE rate of QEs in real time and at room temperature presents a significant challenge as it requires modifying the LDOS at a rate faster than the QEs SE rate (~ 1GHz). The ability to do so would open the door for multiplexing in optical communication and modulation of lasers. Recent works dynamically controlled the lifetime of quantum emitters at cryogenic temperatures by controlling the radiation field in real time [24] or by modifying the exciton-cavity coupling strength [25]. Instead, CET mechanism provides real time control over the SE rate at room temperature through varying the number of QEs participating in cooperative energy transfer to a plasmonic nanoantenna. This is corroborated by demonstrating a reversible dynamic control over the SE rate upon reversibly varying the excitation intensity (Fig. 3c). Regions with white background represent data taken when the excitation intensity was decreased from 37 to 4.4 W. $cm^{-2}$, whereas light blue regions represent data taken by increasing the excitation intensity from 4.4 to 37 W. $cm^{-2}$. This reversible response offers a complete control on the SE rate and establish the basis for a novel class of optical modulators. Note that in the fourth region, the SE rates are slightly lower for all excitation intensities. This is due to QDs bleaching over long exposure times which decreases the number of excited emitters, hence, the cooperative energy transfer rate.

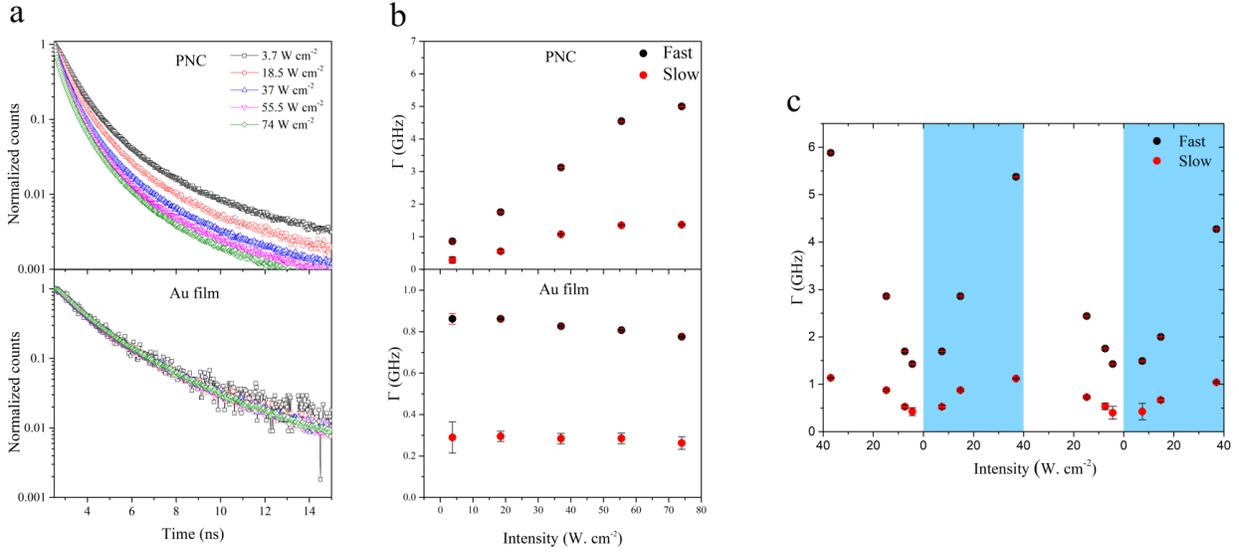

**Figure 3| Intensity dependence of SE rate:** (**a**) Measured time-resolved photoluminescence for five different excitation intensities for the PNC **(Top)** and the reference Au film **(Bottom).** The SE lifetime is intensity dependent for the PNC case only. (**b**) The fitted SE rate fast component (black spheres) and slow component (red spheres) for the PNC **(Top)** and for the reference Au film **(Bottom).** (**c**) Reversible, dynamic control over SE rate. The fast (black dots) and slow (red dots) SE rate components vary by modifying the number of excited emitters. The SE rate is also linearly proportional to the excitation intensity.

To quantitatively demonstrate that the linear dependence of the measured SE rate is due to CET, we first investigate the origin of the fast ($\Gamma^{fast}$) and slow ($\Gamma^{slow}$) SE rates. **Figure 4a** shows the ratio ($\Gamma^{fast}/\Gamma^{slow}$) of QDs on the reference Au film as a function of intensity (black dots) is ~ 3 which suggests that the fast and slow rates correspond to the emission of charged biexcitons and charged excitons, respectively, according to the statistical scaling law at room temperature [31]. This is because a charged biexciton (3 electrons and 2 holes) have six decay pathways via radiative electron-hole recombination, while a charged exciton (2 electrons and 1 hole) has only two decay pathways (**Fig. 4a** inset). As we mentioned above, the SE rate of a QD coupled to a large nanoantenna is approximately equal to the energy transfer rate ($\Gamma^{ET}$). Accordingly, the same statistical scaling applies to energy transfer rates, i.e., $\Gamma_{ET}^{fast}/\Gamma_{ET}^{slow}$ ~ 3. Below the saturation intensity of the QDs, the number of excited QDs participating in CET scales linearly with the excitation intensity $I$ with a scaling factor $\alpha$, i.e., $N = \alpha I$ (since excited QDs' number is an integer, *N* here is understood as its average over some small intensity range). Consequently, the experimentally measured SE rate $\Gamma^{Exp}(I)$ below saturation for QDs participating in CET is given by

$$\Gamma^{Exp}(I) = \Gamma^{ET} + \alpha \, \Gamma^{ET} I \tag{1}$$

where the second term represents the cooperative energy transfer rate in the CET intensity range. Note that for weak excitation intensities when only few emitters are excited, cooperative effects are expected to be small and the experimentally measured SE rate $\Gamma^{Exp}$ should equal the individual QD energy transfer rate $\Gamma^{ET}$. Equation (1) should hold for both the fast and slow rates. Accordingly, the ratio of the experimentally measured fast ($\Gamma^{fast}$) and slow ($\Gamma^{slow}$) rates from the PNC is

$$\Gamma^{fast}(I)/\Gamma^{slow}(I) = (\Gamma_{ET}^{fast} + \alpha_{fast} \, \Gamma_{ET}^{fast} I)/(\Gamma_{ET}^{slow} + \alpha_{slow} \, \Gamma_{ET}^{slow} I) \tag{2}$$

where $\alpha_{fast}$ and $\alpha_{slow}$ are the intensity scaling factors for fast and slow energy transfer rate, respectively. The rates ratio $\Gamma^{fast}(I)/\Gamma^{slow}(I)$ for different intensities is ~3 (red dots), which can only be true if $\alpha_{fast} \approx \alpha_{slow} \approx \alpha$. Importantly, since we have two equations and one unknown, $\alpha$, we can quantitatively validate our analysis by using the measured slow rate $\Gamma^{slow}(I) = \Gamma_{ET}^{slow} + \alpha \, \Gamma_{ET}^{slow} I$, to calculate $\alpha$ which we use to reproduce the experimentally measured fast rate $\Gamma^{fast} = \Gamma_{ET}^{fast} + + \alpha \, \Gamma_{ET}^{fast} I$. **Figure 4b** shows the calculated vs. measured $\Gamma^{fast}$, which are in close agreement, indicating that the slope of SE rate intensity dependence is proportional to the energy transfer rate of individual emitters, as predicted by the proposed CET mechanism. For relatively higher intensities, the rate ratio $\Gamma^{fast}/\Gamma^{slow}$ exceeds 3 likely because excitons saturate at lower intensities compared to biexcitons [32]. We provide the same analysis presented in **Fig. 4a** and **Fig. 4b** for data collected from a different PNC to prove the reproducibility of our observation (*Supplementary information,* **Fig. S5**).

**Figure 4c** shows the photoluminescence from a PNC as a function of excitation intensity. The photoluminescence spectrum clearly retains the plasmon resonance central frequency and overall line-shape while its amplitude increases linearly with excitation power, implying that radiation emanates from the plasmonic antenna following CET from the excited QEs [18]. This is in stark contrast to super-radiance where radiation emanates directly from QEs and so the changes in the decay rates affect accordingly the emission spectra [21]. Furthermore, we unequivocally exclude stimulated emission as a cause of the emission rate intensity dependence by performing photoluminescence and time resolved lifetime measurements as a function of excitation intensity on the same PNC (*Supplementary information,* **Fig. S6**). The lack of stimulated emission was evident due to the absence of nonlinear growth in the emission peak maximum and the complete absence of band narrowing in the photoluminescence spectra. We also ruled out the effect of elevated temperature inside the cavity on the observed lifetime measurements (*Supplementary information*).

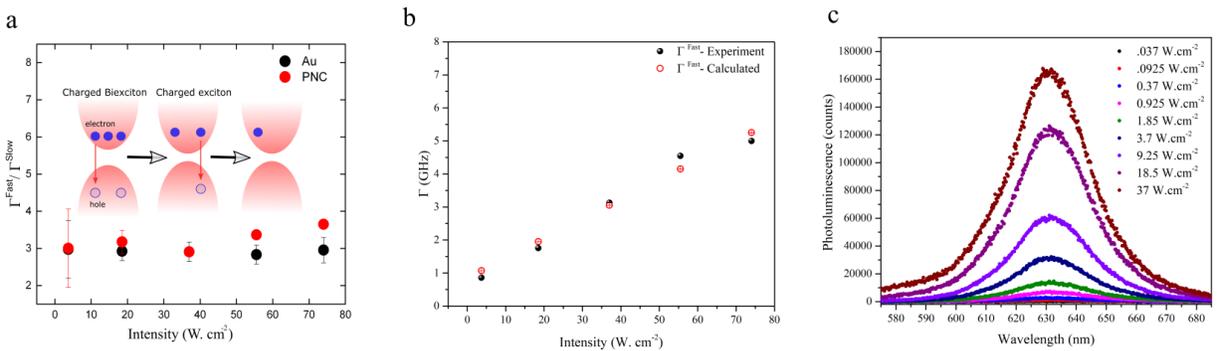

**Figure 4| Analysis of intensity dependent SE rate due to CET: (a)** The ratio of the measured fast $\Gamma^{fast}$ and slow $\Gamma^{slow}$ SE rates for QDs on the reference Au film and inside the PNC. The ratio $\Gamma^{fast}(I)/\Gamma^{slow}(I)$ is ~ 3. Inset: schematic of the decay process of charged biexcitons and charged excitons. **(b)** The rate $\Gamma^{fast}$ is calculated from experimental rate $\Gamma^{slow}$ by assuming that the slope of the SE rate vs. intensity curve is proportional to the energy

transfer rate of individual QD, as predicted by equation (2). (c) The photoluminescence as a function of excitation intensity show that the emission spectrum retains the plasmon lineshape as the peak emission wavelength is ~ 631 nm.

The speeding up of the spontaneous emission based on CET represent an additional degree of freedom to control SE beyond the plasmon-enhanced local field [7]. In this work, we used a low *Q* antenna to increase the antenna radiative losses and ensure that the collected photoluminescence is from the antenna. Future works can use high *Q* and low *V* nano-antennas [4], to enhance the SE rate beyond that of stimulated emission ( > 100 GHz) which would be a landmark in the field of nanophotonics as it will enable short-distance optical communication, and enhance the efficiency of SE based light sources, e.g. Si based light emitting diodes [6,33,34]. The SE rate acceleration of QDs is particularly important as it increases the QDs quantum yield by overcoming Auger recombination. Furthermore, the demonstrated SE rate based optical modulator can be used as a multiplexing technique to encode information in the emission rate. Finally, the versatility of ultrafast phenomena in plasmonics would allow for different strategies to dynamically control the SE rate which can follow our work [35].

**Acknowledgments:** The authors would like to thank Eugenio Calandrini for his help with sample preparation. **Funding:** G.S. received funding from the Ohio Third Frontier Project 'Research Cluster on Surfaces in Advanced Materials (RC-SAM) at Case Western Reserve University' and the GU Malignancies

Program of the Case Comprehensive Cancer Center. FDA received funding from the European Research Council under the European Union's Seventh Framework Programme (FP/2007-2013)/ERC Grant Agreement no. [616213], CoG: Neuro-Plasmonics. J.B. received support from U.S. Department of Energy, Office of Science, Basic Energy Sciences, under Award #DE- SC008148. T.V.S was supported in part by NSF grants No. DMR-1610427 and No. HRD-1547754.

**Author contributions** M.E., T.V.S and G. S. conceived the idea. F.D.A. and E.M. designed and fabricated the samples. M.E., J.B. and G.S. designed the experiments. M.E., A.F., M.W. and E.H. performed experiments. A.B. performed simulations. M.E. wrote the manuscript with inputs from all the authors. T.V.S., J.B., F.D.A., and G.S. supervised the research. All authors analyzed and discussed the data.

# Supplementary information:

# Cooperative Energy Transfer Controls the Spontaneous Emission Rate Beyond Field Enhancement Limits

Mohamed ElKabbash[1*], Ermanno Miele[1,2], Ahmad K. Fumani[1], Michael S. Wolf[1], Angelo Bozzola[2], Elisha Haber[1], Tigran V. Shahbazyan[3], Jesse Berezovsky[1], Francesco De Angelis[2], Giuseppe Strangi[1,2,4*].

[1.] Department of Physics, Case Western Reserve University, 10600 Euclid Avenue, Cleveland, Ohio 44106, USA.

[2.] IIT - Istituto Italiano di Tecnologia, via Morego 30, 16163 Genova, Italy.

[3.] Department of Physics, Jackson State University, Jackson, Mississippi 39217, USA.

[4.] CNR-NANOTEC, Istituto di Nanotecnologia and Department of Physics, University of Calabria, Italy.

*Correspondence authors E-mail: mke23@case.edu, gxs284@case.edu.

## 1- Supplementary Methods:

### 1.1- Materials:

The quantum dots are Lumidot CdSe/ZnS purchased from Sigma-Aldrich (Quantum Dots, QDs, 640 nm emission peak, core-shell type quantum dots, 5 mg/mL in toluene Sigma-Aldrich, Cat No. 680646-2mL).

### 1.2- Fabrication of plasmonic nanocavity:

Out-of-plane plasmonic structures have been fabricated by a focused ion beam assisted nanofabrication technique, described in [1,2]. We slightly modified the fabrication process to incorporate the emitters in the vertical structures. Shipley S1813 has been deposited by spin coating onto a 100 nm SiNx membrane. Resist thickness has been adjusted to achieve the final 450 nm (±50 nm) nano-pillars height. Polymeric cylindrical scaffold (20 nm wall thickness) is defined by ion milling (Helios Nanolab 620, FEI Co.,

Hillsboro, OR, USA) to form cylindrical structures. Acceleration voltage was set to be 30 kV with beam current of 40 pA. CdSe/ZnS nanocrystals have been deposited by spin coating (1000 rpm) after dilution to obtain 0.5 mg/mL solution in toluene. Gold (20 nm) was deposited by DC sputter coating in tilted geometry to obtain a uniform metal mantel deposited onto gain material. QDs ligands (Hexyldecylamine (HDA), trioctylphosphine (TOPO)) acted as spacer in between gold and CdSe/ZnS. The outer and inner diameters of the cavity are approximately 120 and 40 nm, respectively. Scanning Electron Microscopy (SEM, Helios Nanolab 620, FEI Co., Hillsboro, OR, USA) was employed to characterize the morphology of fabricated structures. QDs infiltrated structures were prepared by letting diffuse QDs inside nano-pillar cavity. A droplet (10 μl) of 2.5 mg/ml QDs solution was placed on the backside of the membrane of as-milled nano-pillars. Finally, gold was uniformly deposited onto QDs infiltrated polymeric scaffolds to form 20 nm layer, as previously described.

**1.3- Fabrication of the reference sample:**

The reference sample is prepared by first depositing a 5 nm Ti adhesion layer on a glass slide, then we deposited 45 nm Au layer. QDs were spin-coated with the same concentration and spin coating parameters used for the QDs spin-coated on the main sample.

**1.4- Measuring the LSP resonance of the Plasmonic nanocavity:**

We used a Leica DM2500P microscope and attached an Ocean Optics HR4000CG optical fiber to turn it to a spectrometer. By placing the sample under a 100X objective lens with NA 0.75, we were able to collect the scattered light through the objective off and on the nano-pillar array. By taking the difference in the scattered light we obtain the distinct scattering of the nano-pillars which corresponds to its plasmon resonance.

**1.5- Photoluminescence and fluorescence lifetime measurements:**

Excitation is provided by a supercontinuum fiber laser (Fianium SC450PP), outputting ≈25 ps duration pulses at a repetition rate of 0.2 MHz. The spectrally broad output of the laser is then filtered by a linearly-graded high-pass and low-pass filter (Edmond optics) mounted on motorized translation stages to tune the cut-on and cut-off wavelengths. The excitation wavelength for the QDs was 490nm with a bandwidth of 20 nm corresponding to its full width half maximum. The excitation (pump) beam is passed through short pass filters with cut-off wavelength of 530 nm. After passing through a 50:50 cube, the beam was then focused on the sample via a 100X objective lens (0.75 NA). The photoluminescence (PL) and scattering is collected again via the 100X microscope objective where we place a long pass 550 nm filter to eliminate the excitation beam when necessary. The PL proceeds to an electron-multiplication CCD (EMCCD) camera/spectrometer to either image the PL or the PL spectrum. By placing the sample on a 3D automated translation stage, we could locate the pillars via their PL image. The PL can also be sent to a pair of avalanche photodiodes (APD, PDM-50ct) connected to a time-correlated single photon counting system (TCSPC, Hydraharp 400). The power density was calculated by first measuring the average power before focusing, then determining the beam spot diameter after focusing using the CCD camera. The calculated diameter is ~ 0.6 μm.

### 1.6- Simulating the nano-pillar plasmonic resonance:

All the electromagnetic calculations have been carried out using the commercial software Comsol Multiphysics® (RF module). For the calculations regarding the out-of-plane antennas (Fig. 1d and 1f.), we assumed a TM polarized plane wave with an angle of incidence of 0°. The simulation domain is enclosed within perfectly matched layers (PML) to avoid spurious reflections from the boundaries of the simulation box. The surface is determined by the bottom boundaries of the simulation box and by the metallic bottom-plane

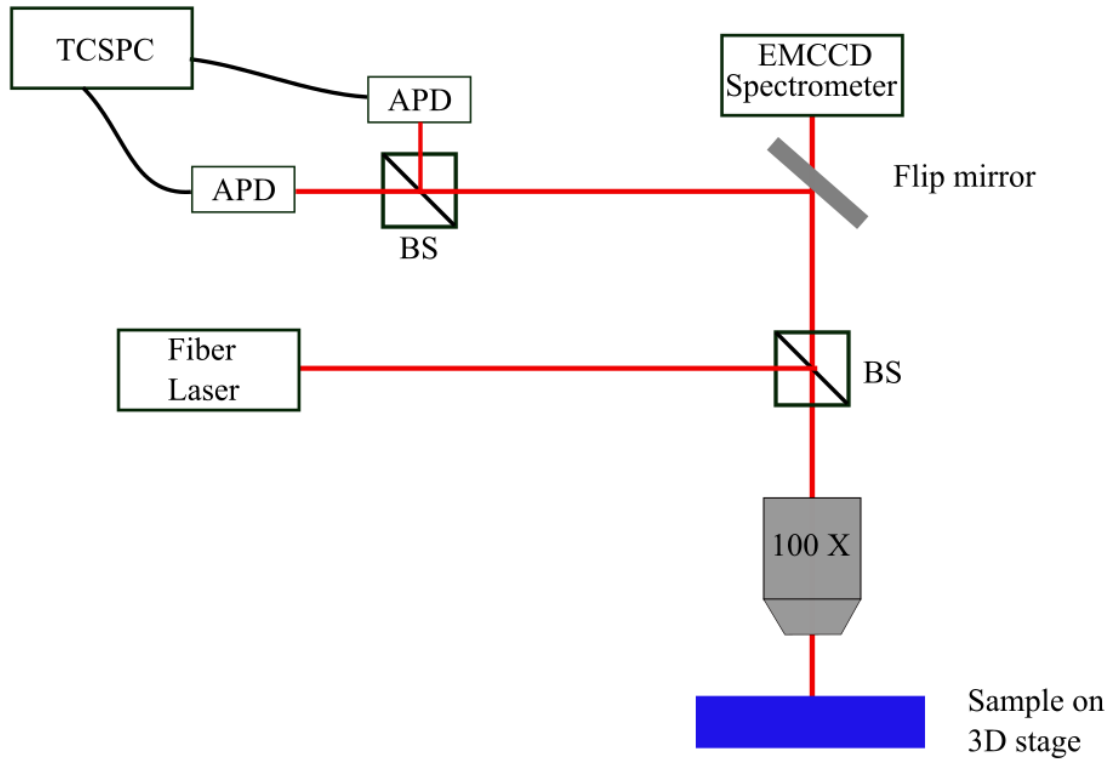

**Figure S1: Schematic of the PL and transient fluorescence spectroscopy setup.** The setup allows us to measure the lifetime and the photoluminescence from the same PNC. In addition, by using an ICCD camera, we can locate the PNC by looking at the QDs emission.

## 2- Supplementary notes:

### 2.1-Radiation pattern from the nanopillar PNC:

To maximize the collected radiation from the nanoantenna, the antenna must be highly radiation, and the radiation must be directional. The nanopillar, out-of-plane, antennas are highly radiative when the nanoantenna height is > 350nm. Figure S2 shows COMSOL simulation of the radiation pattern from the nanoantenna. Clearly, having an antenna with

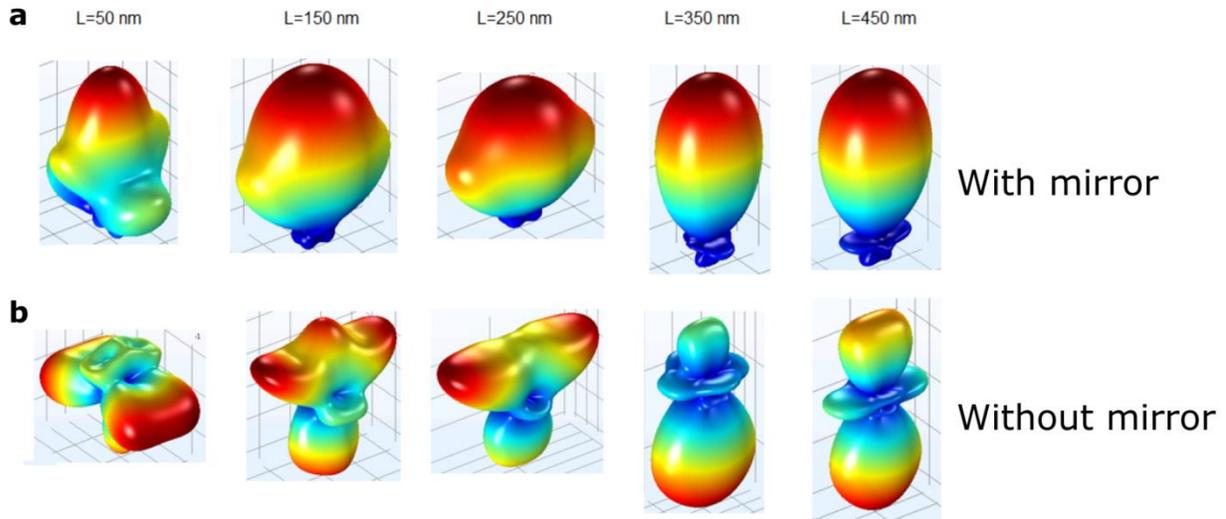

**Figure S2| Directional radiation of out-of-plane nanocavity.** The simulated radiation pattern of a single plasmonic nanocavity (PNC) as a function of the PNC height (L) a) with, and b) without a mirror at its bottom. The height of the PNC determines the directionality of the PNC. Adding a mirror results in unidirectional radiation.

## 2.2- QDs absorption:

As we detailed in the manuscript, following the energy transfer process to the PNC, the energy pathways are radiation to free space, energy dissipation inside the PNC, or energy transfer from the PNC to the QDs. In our experiments, we want to maximize the radiation of the transferred energy to the free space. One major quality of the quantum emitter used, thus, must be to have minimum overlap between its absorption band with the plasmon band to minimize reabsorption. In our case, since we require an emitter to have its emission band with the plasmon band, we need to have an emitter that does not have strong overlap between its absorption and emission band, i.e., one with large Stokes shift. Quantum dots are ideal candidates. Figure S2 shows the absorption spectrum of the used CdSe/ZnS quantum dots. The maximum emission wavelength (638 nm) (noted here by a dotted line) has small overlap with the QDs absorptance.

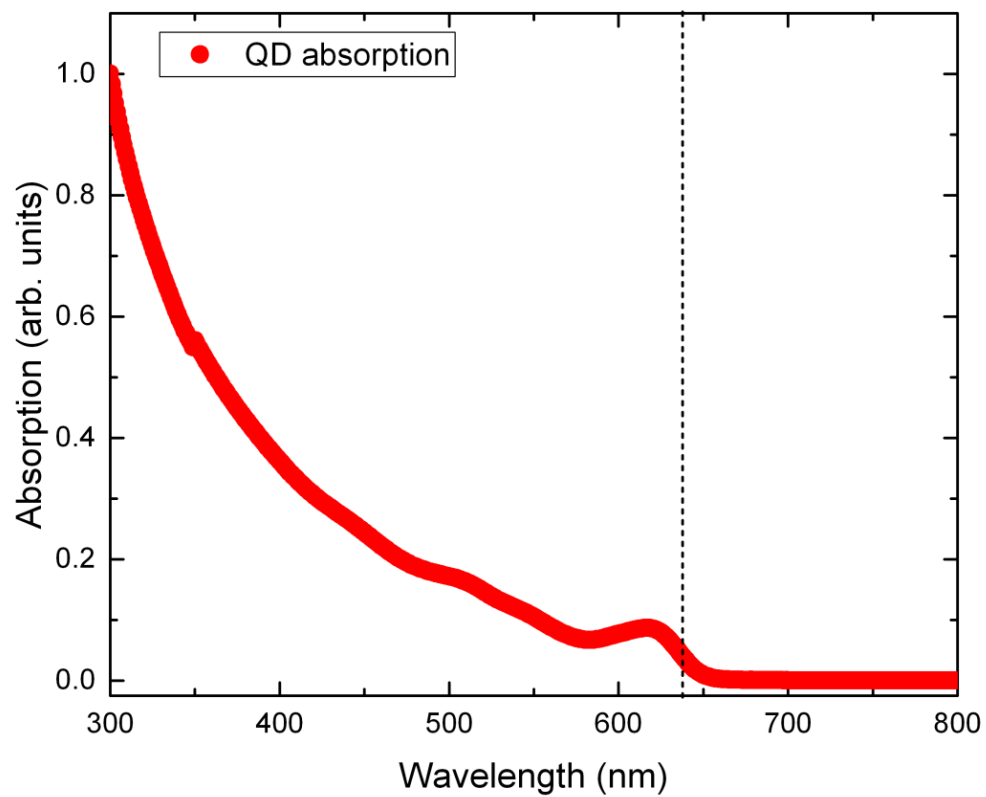

**Figure S3| Absorption of CdSe/ZnS quantum dots**. The absorption of the QDs is maximum in the UV region of the spectrum and significantly decay in the visible. The absorption is weak where the emission of the QDs takes place. This is an important condition to ensure that the energy is channeled mainly towards creating surface plasmons that decay radiatively.

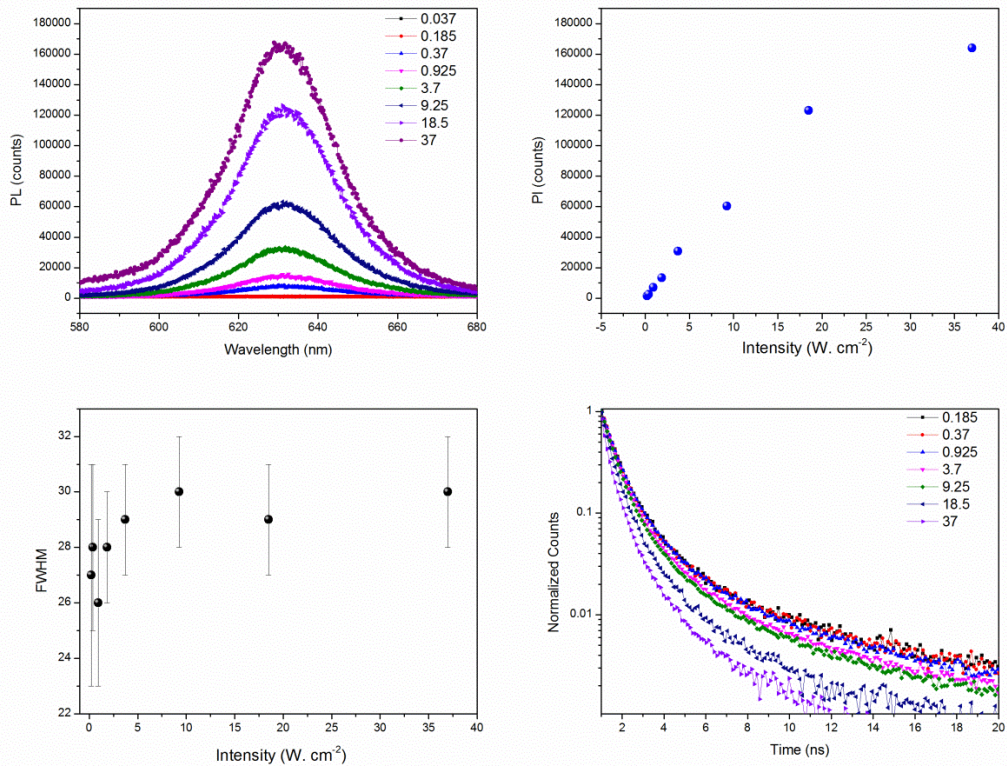

**Figure S3| Excluding stimulated emission and photo-thermal heating. (a)** The PL spectrum for different excitation intensity which shows no amplification due to stimulated emission. This is clearly corroborated by **(b)** the lack of nonlinear growth in the emission peak maximum, and **(c)** the constant FWHM of the PL spectrum. On the other hand, the fluorescence lifetime for the same intensity range and the same PNC shown in **d)** exhibits significant decrease (SE rate increase) as a function of excitation intensity. Note that **(a)** also serves to exclude the effect of photothermal heating in changing the emission rate. Quantum dots are temperature sensors and their PL spectrum experience broadening and quenching, and the peak is red-shifted if their surrounding environment temperature is elevated.

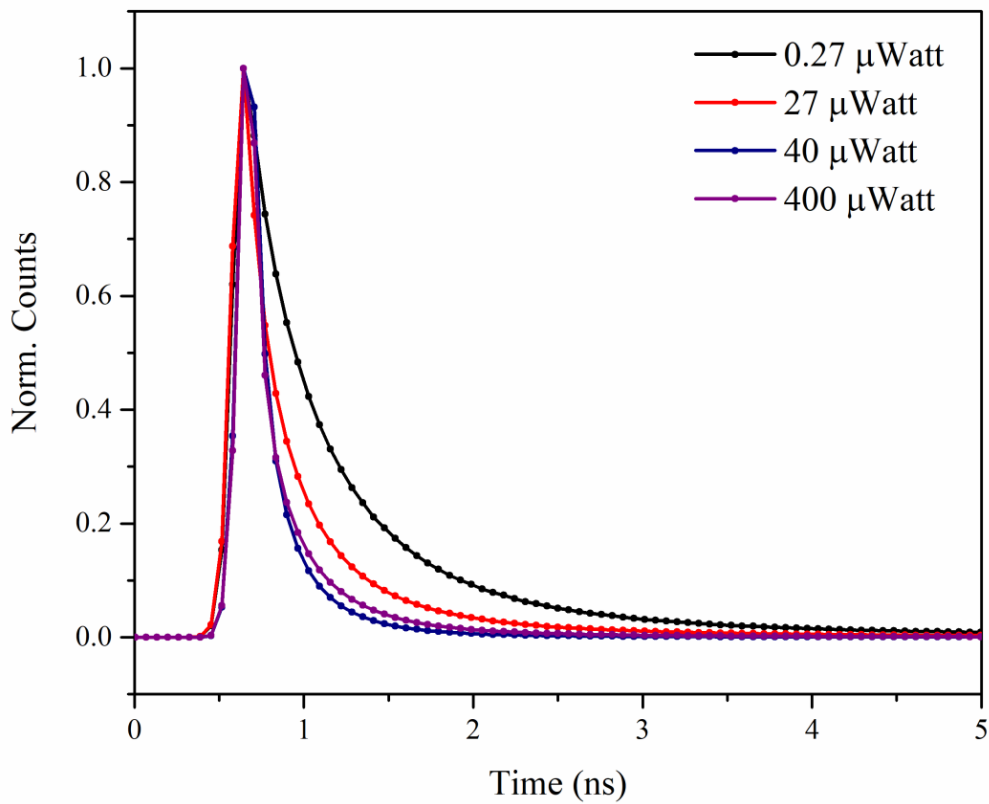

**Figure S4| Effect of photothermal heating on the SE lifetime.** Performing a time fluorescence measurement on a nanopillar infiltrated (not spin coated) with QDs allows us to use high pump intensity and obtain a reasonable signal. The lifetime was measured as we changed the pump power. The lifetime drops as we observed as a function of intensity up to 40 μWatts. However, at 400 μWatt, the lifetime increases. This result is in agreement with previous works[3] on similar QDs and conclusively rule out any effect of thermal heating on the observed SE rate behavior.

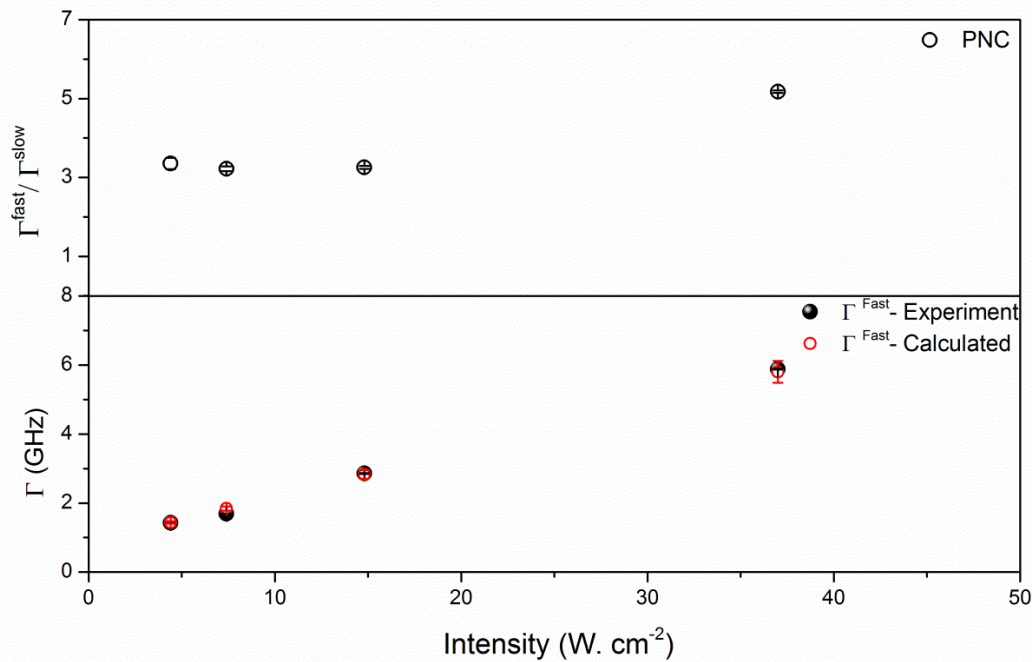

**Figure S5|** The ratio of the measured fast $\Gamma^{fast}$ and slow $\Gamma^{slow}$ SE rates for QDs the PNC presented in Fig. 4 in the main manuscript. The ratio is approximately 3 which indicates that the fast and slow rates are due to the excitation of charged biexcitons and charged excitons, respectively. **b)** By assuming plasmon assisted resonant energy transfer, we calculate $\Gamma^{fast}$ from $\Gamma^{slow}$. The remarkable agreement between the calculated and measured $\Gamma^{fast}$ clearly shows that PCET is the process responsible for the SE rate modification.